\documentclass[journal]{IEEEtran}
\usepackage{graphicx}
\ifCLASSINFOpdf
   \usepackage{graphicx}
\else
 \fi

\usepackage{amsmath, amssymb, amsthm}
\usepackage{algorithm}
\usepackage{algorithmic}
\usepackage{xcolor}
\usepackage{subfigure}
\usepackage{cite}

\newtheorem{definition}{Definition}[section]
\newtheorem{theorem}{Theorem}[section]
\newtheorem{lemma}{Lemma}[section]
\newtheorem{corollary}{Corollary}[section]
\newtheorem{proposition}{Proposition}[section]
\theoremstyle{plain}
\renewcommand\thesubsubsection{\Alph{subsubsection}}
\newcommand*{\QED}{\hfill\ensuremath{\blacksquare}}

\begin{document}
\title{Efficient Two-Dimensional Line Spectrum Estimation Based on Decoupled Atomic Norm Minimization}
\author{Zhe Zhang,~\IEEEmembership{Member,~IEEE,}
		Yue Wang,~\IEEEmembership{Member,~IEEE,}
        and Zhi Tian,~\IEEEmembership{Fellow,~IEEE,}
\thanks{Part of this work was supported by the NSF grant \#CCF-1527396, \#ECCS-1546604, \#AST-1547329 and \#AST-1443858.}
\thanks{Part of this work was presented on the 42th International Conference on Acoustics, Speech, and Signal Processing (ICASSP 2017), New Orleans, LA, USA, March 2017. The original version of this manuscript was presented for peer reviews on March 4, 2017, and subsequent revisions were made to improve this work.}
\thanks{Z. Zhang, Y. Wang and Z. Tian are with Electrical and Computer Engineering Department, George Mason University. }
}

\markboth{}%
{Zhang Z. \MakeLowercase{\textit{et al.}}: Efficient Two-Dimensional Line Spectrum Estimation Based on Decoupled Atomic Norm Minimization}

\maketitle

\begin{abstract}
This paper presents an efficient optimization technique for gridless {2-D} line spectrum estimation, named decoupled atomic norm minimization (D-ANM). The framework of atomic norm minimization (ANM) is considered, which has been successfully applied in 1-D problems to allow super-resolution frequency estimation for correlated sources even when the number of snapshots is highly limited. The state-of-the-art 2-D ANM approach vectorizes the 2-D measurements to their 1-D equivalence, which incurs huge computational cost and may become too costly for practical applications. We develop a novel decoupled approach of 2-D ANM via semi-definite programming (SDP), which introduces a new matrix-form atom set to naturally decouple the joint observations in both dimensions without loss of optimality. Accordingly, the original large-scale 2-D problem is equivalently reformulated via two decoupled one-level Toeplitz matrices, which can be solved by simple 1-D frequency estimation with pairing. Compared with the conventional vectorized approach, the proposed D-ANM technique reduces the computational complexity by several orders of magnitude with respect to the problem size.
It also retains the benefits of ANM in terms of precise signal recovery, small number of required measurements, and robustness to source correlation. The complexity benefits are particularly attractive for large-scale antenna systems such as massive MIMO, radar signal processing and radio astronomy. 
\end{abstract}


\IEEEpeerreviewmaketitle

\section{Introduction}

\IEEEPARstart{T}{wo}-dimensional ({2-D}) line spectrum estimation is an important signal processing problem that has found broad applications, such as {2-D} direction of arrival (DOA) estimation \cite{lee2003low, kedia1997new}, radar signal processing \cite{nion2010tensor} and wireless communications \cite{wang2012low}. As an extension of the widely-studied one-dimensional (1-D) case, {2-D} line spectrum estimation deals with measurements that result from a linear mixture of {2-D} sinusoids, and the goal is to recover these {2-D} sinusoids effectively under certain constraints.

Plenty of work has been done to solve the {2-D} line spectrum estimation problem, often as extensions to 1-D techniques \cite{roy1989esprit, schmidt1986multiple}. Based on sample statistics, various classical super-resolution subspace methods are developed for the {2-D} case, including {2-D} unitary ESPRIT \cite{haardt19952d}, {2-D} MUSIC \cite{hua1993pencil}, matrix enhancement matrix pencil \cite{hua1992estimating}, and tensor-based estimation \cite{nion2010tensor}. These techniques work effectively  provided that multiple snapshots are available to well approximate the signal covariance via sample averaging.  
Moreover, they typically require the sources to be uncorrelated. 
In order to overcome those issues, smoothing techniques have been incorporated into subspace methods, such as 2-D damped MUSIC (DMUSIC) \cite{li1998high}, 2-D RARE \cite{pesavento2004multidimensional} and 2-D IMDF \cite{liu2006eigenvector, liu2006statistical}. With smoothing, these methods work  for correlated sources even with a single snapshot, at the expense of reducing the effective array aperture size and hence sacrificing the resolution. Besides, subspace methods rely on some prior knowledge of the signal, such as the number of sources. 

Alternatively, compressive sensing (CS) is a recent structure-based signal processing framework that suggests one can recover a signal from highly compressed samples if the original signal is sparse under some basis \cite{donoho2006, candes2006, candes2006stable}. When the source signal has a sparse support on the frequency domain, CS may work for line spectrum estimation even with a single snapshot regardless of the signal correlation, without reducing the array aperture size\cite{carlin2013directions, schniter2014channel}. 
Unfortunately, the conventional CS approach suffers from limited resolution and basis mismatch when the true signal frequencies are off-grid \cite{herman2010general, chi2011sensitivity}. This is a huge disadvantage  in performance compared with traditional subspace methods.

Developed as a gridless CS approach, a recent line of work resorts to atomic norm minimization (ANM) and semi-definite programming (SDP) to enable recovery of off-grid sinusoidal components from compressive measurements \cite{bhaskar2013atomic, candes2014towards, tang2013compressed, yang2015gridless}. It exploits the Vandermonde structure of the signal to attain off-the-grid estimation at super-resolution. {Similar results have been developed via total-variation norm minimization when measurements are collected along spectral lines 
\cite{doi:10.1137/16M1108807}}. Encouraged by the successful application in 1-D problems, ANM is also applied to {2-D} spectral estimation problems \cite{xu2014precise, chi2015compressive, yang2016vandermonde}. The main idea is to vectorize the {2-D} signal of interest and then cast the {2-D} Vandermonde  structure of the signal into a proper SDP formulation for ANM. Vectorization-based ANM has shown great performance benefits including single snapshot applicability, resilience to signal correlation and off-grid recovery, but it is highly expensive in  computation because of the huge problem scale resulted from the vectorization operation. Simulations show that on a regular PC, vectorized ANM of a $32\times 32$ {2-D} line spectrum estimation problem cannot be completed in two days. This fact limits the application of {2-D} ANM in practice.

The goal of this paper is to develop a new formulation for {2-D} ANM that retains the performance benefits of vectorized ANM while remarkably reduces its computational complexity. 
By introducing a new matrix-form atom set, we present a novel decoupled approach of {2-D} ANM via SDP to naturally decouple the joint observations in both dimensions without loss of optimality. Without any relaxation or approximation, the proposed decoupled-ANM (D-ANM) strategy reformulates the original large-scale {2-D} problem into a reduced-size formula expressed by two decoupled one-level Toeplitz matrices, which leads to simple 1-D frequency estimation with pairing. Compared with the existing vectorized approach, our proposed technique reduces the computational complexity from the order of $\mathcal{O}(N^{3.5}M^{3.5})$ to $\mathcal{O}((N+M)^{3.5})$, where $N$ and $M$ are the problem sizes of the two dimensions respectively. This is a dramatic acceleration in computational efficiency. 

The rest of this paper is organized as follows. In Section \ref{sec:bak}, the problem formulation is introduced. The vectorization-based 2-D ANM approach is reviewed in Section \ref{sec:sub}. 
Main results of decoupling via SDP are developed in Section \ref{sec:res}, followed by the complete D-ANM formulation in Section \ref{sec:danm}. Further issues, such as compression, resolution and complexity are discussed in Section \ref{sec:dis}. Numerical results are given in Section \ref{sec:sim} to validate the proposed D-ANM method, followed by conclusions in Section \ref{sec:con}.

\section{Signal Model and Problem Statement}
\label{sec:bak}

Consider a {2-D} line spectrum estimation problem where the signal of interest $\mathbf{X}(t)\in\mathbb{C}^{N\times M}$ is a linear mixture of $L$ {2-D} sinusoidal components in the form of
\begin{equation}
	\label{eq:2.1}
	\mathbf{X}(t)=\sum_{l=1}^{L} s_l(t) \mathbf{a}_N(f_{x, l})\mathbf{a}_M^\mathrm{H}(f_{y, l})=\sum_{l=1}^{L} s_l(t) \mathbf{A}(\mathbf{f}_l)
\end{equation}
where $s_l(t)$ is the complex amplitude of the $l$-th 2-D sinusoid at the time snapshot $t, t=1, \dots, T$, $L$ is the number of sources, and $f_{x, l}$ and $f_{y, l}$ are digital frequencies along two orthogonal dimensions respectively, with $\mathbf{f}_l=(f_{x, l}, f_{y, l})^\mathrm{T}\in[0, 1]^2, \forall l$. The manifold vectors $\mathbf{a}_N(f)\in\mathbb{C}^N$ and $\mathbf{a}_M(f)\in\mathbb{C}^M$ exhibit Vandermonde structures of size $N$ and $M$ respectively, as follows:
\begin{displaymath}
	\mathbf{a}_N(f)=\left(1, \exp(j2\pi f), \dots, \exp(j\pi(N-1)f) \right)^\mathrm{T}
\end{displaymath}
\begin{displaymath}
	\mathbf{a}_M(f)=\left(1, \exp(j2\pi f), \dots, \exp(j\pi(M-1)f) \right)^\mathrm{T}.
\end{displaymath}
In (\ref{eq:2.1}), parameters $\{s_l\}_l$, $\{(f_{x, l}, f_{y, l})\}_l$ and $L$ are all unknown.

{
Note that some signal sources may have overlapping frequencies along one dimension, while being distinct on the other dimension. That is, $\exists i\neq j\in[1, L]$, such that $\mathbf{f}_i \neq \mathbf{f}_j$, but $f_{x, i}=f_{x, j}$ or $f_{y, i}=f_{y, j}$. In this case, we let $L_x$ and $L_y$ denote the number of distinct frequencies along the two dimensions respectively, and let $\mathbf{f}_x \in [0,1]^{L_x}$ and $\mathbf{f}_y\in[0,1]^{L_y}$ denote the respective vectors of distinct frequencies along each dimension. 
%
%
Define the corresponding manifold matrices
\begin{displaymath}
	\mathbf{A}_N(\mathbf{f}_x)=(\mathbf{a}_N(f_{x, 1}), \mathbf{a}_N(f_{x, 2}), \dots, \mathbf{a}_N(f_{x, L_x}))\in\mathbb{C}^{N, L_x},
\end{displaymath}
\begin{displaymath}
	\mathbf{A}_M(\mathbf{f}_y)=(\mathbf{a}_M(f_{y, 1}), \mathbf{a}_M(f_{y, 2}), \dots, \mathbf{a}_N(f_{x, L_y}))\in\mathbb{C}^{M, L_y}.
\end{displaymath}
Then, the signal model (\ref{eq:2.1}) can be rewritten concisely as 
\begin{equation}
	\label{eq:2.2}
	\mathbf{X}(t)=\mathbf{A}_N(\mathbf{f}_x)\mathbf{S}(t)\mathbf{A}_M^\mathrm{H}(\mathbf{f}_y)
\end{equation}
where $\mathbf{S}(t) \in \mathbb{C}^{L_x \times L_y}$ with $\{s_{l}(t)\}_{l=1}^L$ in its elements. If there is no frequency overlapping on any dimension, then $\mathbf{S}(t)=\mathrm{diag}(s_1(t), \dots, s_L(t))$ is diagonal.  Otherwise, it may have off-diagonal elements. 
}

In many applications, the signal $\mathbf{X}(t)$ is not directly observed but over a linear (compressive) measurement operator $\mathcal{C}(\cdot)$. Inflicted with an additive noise $\mathbf{W}(t)$, the measurement  $\mathbf{Y}(t)$ is given by
\begin{equation}
	\label{eq:2.7}
	\mathbf{Y}(t)=\mathcal{C}(\mathbf{X}(t))+\mathbf{W}(t).
\end{equation}

We focus on the single measurement vector (SMV) case with $T=1$, and hence drop the index $t$ in (\ref{eq:2.7}). The goal of 2-D line spectrum estimation is to recover those sinusoidal components, especially the frequency pairs $(f_{x, l}, f_{y, l})$, from the measurements $\mathbf{Y}$. Such a problem arises in many applications concerning frequency analysis or DOA estimation, where the measurement operator in (\ref{eq:2.7}) may vary. We give two examples.

\subsubsection{2-D DOA Estimation}

Let $\mathcal{C}(\cdot)$ be a fully observable system, i.e. no compression applied. For example, in 2-D DOA estimation with uniform rectangle array (URA), the received signal obeys the Vandermonde structure in both dimension as
\begin{equation}
	\mathbf{Y}=\mathbf{X}+\mathbf{W}, 
\end{equation}
which is a variation of (\ref{eq:2.7}).
The goal is to estimate the frequencies $(\mathbf{f}_x, \mathbf{f}_y)$ from the noisy measurements $\mathbf{Y}$ \cite{icassp}.

\subsubsection{Channel Estimation}

In a (massive) MIMO communication system, an $N$-element uniform linear array (ULA) and an $M$-element ULA are employed at the transmitter and the receiver  respectively. The MIMO channel can be modeled as the superposition of $L$ directional channel paths \cite{1031867, 951559, el2012low}:
\begin{equation}
	\mathbf{X}=\sum_{l=1}^{L} s_l \mathbf{a}_N(f_{x, l})\mathbf{a}_M^\mathrm{H}(f_{y, l}),
\end{equation}
where $f_{x, l}=\sin\theta_{r, l}$ and $f_{y, l}=\sin\theta_{t, l}$ denote the angle of arrival (AoA) and the angle of departure (AoD) of the $l$-th path, respectively. For data-aided channel estimation,  
a block of pilot symbols  $\mathbf{C}\in\mathbb{C}^{M\times B}$ are transmitted over $B$ symbol periods, yielding
\begin{equation}
	\label{eq:2.9}
	\mathbf{Y}=\mathbf{X} \mathbf{C}+\mathbf{W},
\end{equation}
where $\mathbf{Y}\in\mathbb{C}^{N\times B}$. 
Obviously, (\ref{eq:2.9}) is a variation of (\ref{eq:2.7}). The goal is to estimate $\mathbf{X}$ along with its structure given $\mathbf{C}$ and $\mathbf{Y}$, which boils down to {2-D} line spectrum estimation \cite{icc}.

\section{Atomic Norm Minimization for 2-D Line Spectrum Estimation}
\label{sec:sub}

\subsection{The ANM principle}

The basic idea of ANM is to express the signal of interest as a (concise) linear combination of a few simple atoms over a known atom set, and the structural information of the atoms is utilized for signal reconstruction from (compressive) measurements.

Suppose that a general signal $\boldsymbol{\Theta}$ is composed of several components. Although the composition is unknown, it is known that these components are of the same structure and belong to a known atom set $\mathcal{A}$ that may have an infinite size. That is,
\begin{equation}
	\label{eq:an0}
	\boldsymbol{\Theta}=\sum_l s_l \mathbf{A}_l, \qquad\mathbf{A}_l\in\mathcal{A}.
\end{equation}
Note that a given signal $\boldsymbol{\Theta}$ might have more than one possible decompositions over the atom set $\mathcal{A}$.

The atomic norm of $\boldsymbol{\Theta}$ over the atom set $\mathcal{A}$ is defined as
\begin{equation}
	\label{eq:an}
	\|\boldsymbol{\Theta}\|_\mathcal{A}=\inf\left\{\sum_l |s_l| \bigg | \boldsymbol{\Theta}=\sum_l s_l \mathbf{A}_l, ~~\mathbf{A}_l\in\mathcal{A} \right\},
\end{equation}
which basically seeks the sparsest (under the $\ell_1$-norm measure) decomposition of $\boldsymbol{\Theta}$ over $\mathcal{A}$. 

{
\begin{definition}
 A signal $\boldsymbol{\Theta}$ is said to be \textbf{sparse} over the atom set $\mathcal{A}$, if $\boldsymbol{\Theta}$ is composed of a small number of atoms in $\mathcal{A}$ through a linear combination as follows: 
	\begin{displaymath}
		\boldsymbol{\Theta}=\sum_{l=1}^{L} s_l \mathbf{A}_l, \qquad\mathbf{A}_l\in\mathcal{A}, \qquad L\ll NM.
	\end{displaymath}
	In other words, $\boldsymbol{\Theta}$ has a sparse representation over the set $\mathcal{A}$.
\end{definition}

}

{Our goal is to retrieve the right hand side of (\ref{eq:an0}) given $\mathbf{X}$.} When $\boldsymbol{\Theta}$ is known \emph{a priori} to have a sparse support over $\mathcal{A}$, it is possible to retrieve its components via the following $\ell_1$-norm minimization:
\begin{equation}
	\label{eq:an1}
	\arg \min_{\{\mathbf{A}_l, s_l\}_l}\left\{\sum_l |s_l| \qquad \mathrm{s.t.~} \boldsymbol{\Theta}=\sum_l s_l \mathbf{A}_l,  ~~\mathbf{A}_l\in\mathcal{A} \right\}.
\end{equation}
{We observe that (\ref{eq:an1}) can be solved via finding the atomic norm $\|\boldsymbol{\Theta}\|_\mathcal{A}$, which is why finding the atomic norm for some proper atom set results in line spectrum estimation.} Note that the calculation of atomic norm is usually very difficult, particularly when the atom set is of infinite size. For some atom sets of special structures, computationally efficient calculation of $\|\boldsymbol{\Theta}\|_\mathcal{A}$ may arise, which we will discuss later.

If the measurement
	$\boldsymbol{\Phi}=\mathcal{C}(\boldsymbol{\Theta})+\mathbf{W}$
is observed from the true signal $\boldsymbol{\Theta}$ through a system operator $\mathcal{C}(\cdot)$ and noise $\mathbf{W}$, then a measurement constraint should be added to the calculation of atomic norm:
\begin{equation}
	\label{eq:an3}
	\min_{\boldsymbol{\Theta}} \|\boldsymbol{\Theta}\|_\mathcal{A}, \qquad \mathrm{s.t.~} \|\boldsymbol{\Phi}-\mathcal{C}(\boldsymbol{\Theta})\|\leq\epsilon,
\end{equation} 
where $\epsilon$ is the noise threshold.
and $\|\cdot \|$ is some proper norm.  
The problem in (\ref{eq:an3}) is termed as \emph{atomic norm minimization} (ANM), which is a convex optimization problem and can be solved by some regularization method.

Obviously, if the atom set is composed of 2-D sinusoids of all possible frequencies and $\mathbf{X}$ as in (\ref{eq:2.1}) is known \emph{a priori} to have a sparse frequency support, it is possible to solve the line spectrum estimation problem via ANM. 
The following sections \ref{sec:pw} and \ref{sec:res} focus on solving ANM in a computationally feasible manner for 2-D line spectrum estimation.

\subsection{Vectorization-based 2D ANM}
\label{sec:pw}

Here we review prior work on solving the 2-D line spectrum estimation problem via vectorization-based ANM \cite{xu2014precise, chi2015compressive, yang2016vandermonde}.


Using the Kronecker product $\otimes$,  the signal $\mathbf{X}$ in (\ref{eq:2.1}) can be vectorized as
\begin{equation}
	\label{eq:2.4}
	\mathbf{x}=\mathrm{vec}(\mathbf{X})=\sum_{l=1}^{L} s_l \mathbf{a}_M^\ast(f_{y, l})\otimes\mathbf{a}_N(f_{x, l})=\sum_{l=1}^{L} s_l \mathbf{a}(\mathbf{f}_l)
\end{equation} 
where 
$\mathbf{a}(\mathbf{f})=\mathbf{a}_M^\ast(f_{y})\otimes\mathbf{a}_N(f_{x})$ is the vectorized manifold vector of length $NM$, for $\mathbf{f}=(f_{x}, f_{y})$.  

It is straightforward to define a vector-form atom set as
\begin{equation}
	\mathcal{A}_V=\{ \mathbf{a}(\mathbf{f}), \quad \forall \mathbf{f}\in[0, 1]\times[0, 1]\},
\end{equation}
whose atomic norm $\|\cdot\|_{\mathcal{A}_V}$ is given by (\ref{eq:an}) accordingly.

It has been shown in \cite{chi2015compressive} { that for a line spectrum estimation problem with few components, if the frequencies are adequately separated to meet the separation condition in \cite{chi2015compressive}[Theorem 1] therein}, then the atomic decomposition in (\ref{eq:2.4}) is guaranteed to be the sparsest one, i.e.,
\begin{equation} \label{eq:2.4b}
	\|\mathbf{x}\|_{\mathcal{A}_V}=\sum_{l=1}^{L} |s_l|,
\end{equation}
where $s_l$ happen to be the coefficients in (\ref{eq:2.4}).

Further, {according to Theorem 1 and Proposition 2 in \cite{chi2015compressive}, if $L\leq\min\{N, M\}$ in addition to the separation condition}, $\|\mathbf{x}\|_{\mathcal{A}_V}$ can be calculated via SDP as follows:
\begin{equation}
	\label{eq:2.6}
	\begin{split}
		\|\mathbf{x}\|_{\mathcal{A}_V}&=\min_{\mathbf{u}, v}\left\{\frac{1}{2}\left(v+\mathrm{trace}\big(\mathbf{T}_\mathrm{2D}(\mathbf{u})\big)\right)\right\}  \\
		\mathrm{s.t.~}\quad& \left(\begin{array}{cc}
		v & \mathbf{x}^\mathrm{H} \\
		\mathbf{x} & \mathbf{T}_\mathrm{2D}(\mathbf{u})
		\end{array}\right)\succeq \mathbf{0}
	\end{split}
\end{equation}
where {$\mathbf{T}_\mathrm{2D}(\mathbf{u})\in\mathbb{C}^{NM\times NM}$ is a two-level Toeplitz matrix with $\mathbf{u}\in\mathbb{C}^{NM}$ being its first row, as defined in \cite{yang2016vandermonde}}.
{The SDP yields $\mathbf{u}$ and hence the two-level Toeplitz matrix $\mathbf{T}_\mathrm{2D}(\mathbf{u})$, in which the true frequencies of interest $(\mathbf{f}_x, \mathbf{f}_y)$ are coded. Mature techniques for two-level Vandermonde decomposition can be used to recover $(\mathbf{f}_x, \mathbf{f}_y)$ from $\mathbf{T}_\mathrm{2D}(\mathbf{u})$ \cite{yang2016vandermonde}.}

The vectorization-based method is a general approach for higher ($\geq 2$) dimensional line spectrum estimation \cite{yang2016vandermonde}. However, a main disadvantage is its high computational complexity because of the multi-level Toeplitz matrix involved. Note that the computational complexity of SDP is determined by the size of the positive semidefinite (PSD) matrix in its constraint. For an $N\times M$ 2-D line spectrum estimation problem, the PSD matrix in (\ref{eq:2.6}) is of  $(NM+1)\times(NM+1)$, which grows rapidly with respect to the problem scale.

\subsection{{Suboptimally-decoupled 1D ANM} }
\label{sect:suboptimal}

{ 
In order to avoid a multi-dimensional search for spectral peaks, a classic approach to decoupling  is to estimate ${\mathbf{f}}_x$ and ${\mathbf{f}}_y$ separately from two 1-D problems \cite{68174}. Along this line, we note in Appendix \ref{sec:app2.2} that our 2-D SMV problem can be viewed as two 1-D MMV problems and solved by two separate SDP formulas (see Lemma A.2). However, such decoupling not only suffers from suboptimal accuracy, but also incurs two SDP solvers. Further, this suboptimal decoupling approach typically requires complex frequency pairing, and does not fully exploit the measurement structure \cite{68174}. 
}

Specifically, we may be able to treat the 2-D SMV measurement as 1-D multiple measurement vector (MMV) measurements in both dimensions as
\begin{equation}
\mathbf{X}=\mathbf{A}_N(\mathbf{f}_x)\mathbf{S}\mathbf{A}_M^\mathrm{H}(\mathbf{f}_y)=\mathbf{A}_N(\mathbf{f}_x)\mathbf{S}_y,
\end{equation}
and
\begin{equation}
\mathbf{X}^\mathrm{H}=\mathbf{A}_M(\mathbf{f}_y)\mathbf{S}\mathbf{A}_N^\mathrm{H}(\mathbf{f}_x)=\mathbf{A}_M(\mathbf{f}_y)\mathbf{S}_x,
\end{equation}
where $\mathbf{S}_y=\mathbf{S}\mathbf{A}_M^\mathrm{H}(\mathbf{f}_y)$ and $\mathbf{S}_x=\mathbf{S}\mathbf{A}_N^\mathrm{H}(\mathbf{f}_x)$ denote the equivalent MMV measurement data. This turns out to be two 1-D MMV harmonic retrieval problems and can be solved by a suitable 1-D method such as 1-D MMV ANM \cite{yang2014exact}. 

By defining the MMV atomic norms as in Appendix \ref{sec:app2.2}, the following SDPs can be calculated,
\begin{equation}
	\begin{split}
		\|\mathbf{X}\|_{\mathcal{A}_x}=&\min_{\mathbf{V}_x, \mathbf{u}_x} \left\{\frac{1}{2\sqrt{N}}\bigg(\mathrm{trace}(\mathbf{V}_x)+\mathrm{trace}\big(\mathbf{T}(\hat{\mathbf{u}}_x)\big)\bigg)\right. \\
		&  ~~~~ \mathrm{s.t.}~ \left(\begin{array}{cc}
		\mathbf{V_x} & {\mathbf{X}}^\mathrm{H} \\
		{\mathbf{X}} & \mathbf{T}({\mathbf{u}}_x)
		\end{array}\right)\succeq \mathbf{0},
	\end{split}
\end{equation}
and
\begin{equation}
	\begin{split}
		\|\mathbf{X}^\mathrm{H}\|_{\mathcal{A}_y}=&\min_{\mathbf{V}_y, \mathbf{u}_y} \left\{\frac{1}{2\sqrt{M}}\bigg(\mathrm{trace}(\mathbf{V}_y)+\mathrm{trace}\big(\mathbf{T}(\hat{\mathbf{u}}_y)\big)\bigg)\right. \\
		&  ~~~~ \mathrm{s.t.}~ \left(\begin{array}{cc}
		\mathbf{V}_y & {\mathbf{X}} \\
		{\mathbf{X}}^\mathrm{H} & \mathbf{T}({\mathbf{u}}_y)
		\end{array}\right)\succeq \mathbf{0},
	\end{split}
\end{equation}
where $\mathbf{V}_x\in\mathbb{C}^{M\times M}$ and $\mathbf{V}_y\in\mathbb{C}^{N\times N}$. The frequencies $\mathbf{f}_x$ and $\mathbf{f}_y$ can be recovered from Toeplitz matrices $\mathbf{T}(\mathbf{u}_x)$ and $\mathbf{T}(\mathbf{u}_y)$ via Vandermonde decomposition, where $\mathbf{u}_x$ and $\mathbf{u}_y$ are first rows of the matrices respectively.

However, this is a suboptimal approach, in which the joint 2-D problem is degenerated to two 1-D problems. While the complexity is reduced remarkably, this approach could cause significant performance degradation because the joint information of the two coupled dimensions is overlooked. 

\section{Matrix-From Atomic Norm and Decoupled SDP}
\label{sec:res}

We propose a novel method that decouples the 2-D frequency information into two separate dimensions to reduce complexity, and at the same time retains the performance optimality by jointly utilizing all information on both dimensions.

Recall the signal model (\ref{eq:2.1})
\begin{equation}
	\label{eq:3.0}
	\mathbf{X}=\sum_{l=1}^{L} s_l \mathbf{A}(\mathbf{f}_l).
\end{equation}
Different from the vectorized ANM,  we introduce a new atom set $\mathcal{A}_M$ as
\begin{equation}
	\label{eq:3.1}
	\begin{split}
			\mathcal{A}_M&=\{ \mathbf{A}(\mathbf{f}), \quad \forall \mathbf{f}\in[0, 1]\times[0, 1]\}\\
			&=\{ \mathbf{a}_N(f_x)\mathbf{a}_M^\mathrm{H}(f_y), \quad \forall f_x\in[0, 1], f_y\in[0, 1]\}.
	\end{split}
\end{equation}
This is a \emph{matrix-form atom set}, which naturally results in a matrix-form atomic norm as
\begin{equation}
	\label{eq:3.2}
	\|\mathbf{X}\|_{\mathcal{A}_M}=\inf\left\{\sum_k |s_k| \bigg | \sum_k s_k \mathbf{A}(\mathbf{f}_k), ~~\mathbf{A}(\mathbf{f}_k)\in\mathcal{A}_M  \right\}.
\end{equation}

Note that the matrix-form atom set is composed of rank-one matrices, and hence (\ref{eq:3.2}) amounts to the atomic norm of low-rank matrices. Since the operator $\mathrm{vec}(\cdot)$ is a one-to-one mapping and $\mathcal{A}_M \leftrightarrow \mathcal{A}_V$ is also a one-to-one mapping, it is straightforward to conclude the following proposition.
\begin{proposition}
	\label{prop:1}
	For  $\mathbf{x}=\mathrm{vec}(\mathbf{X})$ as in (\ref{eq:2.4}) and (\ref{eq:3.0}), it holds that 
	\begin{equation}
		\|\mathbf{X}\|_{\mathcal{A}_M}=\|\mathbf{x}\|_{\mathcal{A}_V}. \label{eq:3.2p}
	\end{equation}
\end{proposition}

{
Next, we develop our main results in three steps:
\begin{enumerate}
	\item Under certain conditions, the atomic decomposition of $\mathbf{X}$ over $\mathcal{A}_M$ yielding the atomic norm $\|\mathbf{X}\|_{\mathcal{A}_M}$ is unique, and turns out to be (\ref{eq:3.0}) for the given matrix $\mathbf{X}$. 
	\item $\|\mathbf{X}\|_{\mathcal{A}_M}$ can be efficiently calculated via SDP in a decoupled manner, possibly under stronger conditions. 
	\item The desired frequency pairs $(\mathbf{f}_x, \mathbf{f}_y)$ can be retrieved from the output of SDP.
\end{enumerate}}
The feasibility conditions in these steps concern the frequency separation of sinusoids, indicated by  $|f_{x, i}-f_{x, j}|$ and $|f_{y, i}-f_{y, j}|$, $\forall i\neq j$. 

\subsection{Uniqueness of Atomic Decomposition}

The first step is to assess the \emph{uniqueness} of true signal frequency set as the solution producing the atomic norm. 
The result is given in the following theorem.
\begin{theorem}
	\label{th:1}
	Consider an $N\times M$ data matrix $\mathbf{X}$ given by (\ref{eq:3.0}). 
	{If the frequency components of  $\mathbf{X}$ are adequately separated}\footnote{{The frequency separation condition herein is the same as that for vectorized ANM, as specified by Eq. (10) in Theorem 1 of \cite{chi2015compressive}. The detail is omitted here, since this work will eventually require a stronger separation condition (\ref{eq:2.5}) in Theorem \ref{th:2} which meets this condition as well. On the other hand, since we are not concerned with missing entries in the data matrix $\mathbf{X}$, Theorem \ref{th:1} guarantees the exact and unique recovery of $\mathbf{X}$, but \cite{chi2015compressive} states a probabilistic guarantee due to random missing entries in the data. Though not explicitly stated in \cite{chi2015compressive}, the results and proof for vectorized ANM under no missing entries can be found in \cite{chi2015compressive}[Appendix B, Proof of Theorem 1].}}, then it is guaranteed that (\ref{eq:3.0}) is the unique sparsest atomic decomposition of the data $\mathbf{X}$, yielding  
	\begin{equation} \label{eq:th1}
		 \|\mathbf{X}\|_{\mathcal{A}_M}=\textstyle \sum_l |s_l|
	\end{equation}
	where  $s_l$ are the coefficients in (\ref{eq:3.0}).
\end{theorem}

The proof follows directly from (\ref{eq:2.4}), (\ref{eq:2.4b}), (\ref{eq:3.2}) and (\ref{eq:3.2p}). A complete proof is given in Appendix \ref{sec:app1}. 
{Theorem \ref{th:1} ensures that if the signal of interest is composed of adequately separated sinusoids, its component atoms can be uniquely identified  via finding its atomic norm (\ref{eq:3.2}). } 
On the other hand, 
calculating the atomic norm (\ref{eq:3.2}) is an infinite programming problem over all feasible $\mathbf{f}$, which is difficult. 

\subsection{Calculation of Atomic Norm via Decoupled SDP}

The second step is to reformulate the problem of atomic norm calculation using SDP, for computational efficiency. To do so, a stronger frequency separation condition is invoked, which hinges on the following frequency separation quantities:
\begin{equation}
		\Delta_{\min, x}=\min_{i\neq j} |f_{x, i}-f_{x, j}|, \quad
		\Delta_{\min, y}=\min_{i\neq j} |f_{y, i}-f_{y, j}|.
\end{equation}
 
The following theorem arises.
\begin{theorem}
	\label{th:2}
	If the following \emph{sufficient frequency separation condition} holds:
	\begin{equation}
		\label{eq:2.5}
		\Delta_{\min, x}\geq\frac{1}{\lfloor(N-1)/4\rfloor}, \mbox{~~or~~}\Delta_{\min, y}\geq\frac{1}{\lfloor(M-1)/4\rfloor},
	\end{equation}
  and {$L\leq \min\{M,N\}$,} then the matrix-form atomic norm in (\ref{eq:3.2}) can be efficiently computed via the following SDP:
	\begin{equation}
		\label{eq:3.3}
		\begin{split}
			\|\mathbf{X}\|_{\mathcal{A}_M}&=\min_{\mathbf{u}_x, \mathbf{u}_y} \left\{\frac{1}{2\sqrt{NM}}\bigg(\mathrm{trace}\big(\mathbf{T}(\mathbf{u}_x)\big)+\mathrm{trace}\big(\mathbf{T}(\mathbf{u}_y)\big)\bigg)\right\}  \\
			& \mathrm{s.t.}\quad \left(\begin{array}{cc}
			\mathbf{T}(\mathbf{u}_y) & \mathbf{X}^\mathrm{H} \\
			\mathbf{X} & \mathbf{T}(\mathbf{u}_x)
			\end{array}\right)\succeq \mathbf{0}
		\end{split}
	\end{equation}
	where $\mathbf{T}(\mathbf{u}_x)\in\mathbb{C}^{N\times N}
	$ and $\mathbf{T}(\mathbf{u}_y)\in\mathbb{C}^{M\times M}$ are one-level Hermitian Toeplitz matrices defined by the first rows $\mathbf{u}_x\in\mathbb{C}^N$ and $\mathbf{u}_y\in\mathbb{C}^M$ respectively. 
\end{theorem}

The proofs of Theorem \ref{th:2} can be found in Appendices \ref{sec:app2}. 

{

\subsection{Frequency Identification}

Finally, we show that the desired frequency pairs can be indeed retrieved from the SDP.
\begin{corollary}
	\label{co:1}
	{The two Toeplitz matrices $\mathbf{T}(\mathbf{u}_x^\star)$ and $\mathbf{T}(\mathbf{u}_y^{\star})$ in (\ref{eq:3.3}) are both positive semidefinite and low rank, of rank $L_x$ and $L_y$ respectively, whose one-level Vandermonde decomposition corresponds to the true signal frequencies $\mathbf{f}_x$ and $\mathbf{f}_y$ respectively. }
\end{corollary}

This corollary is straightforward from the proof of Lemma \ref{lm:a1} in Appendix \ref{sec:app2}. }

{Theorem \ref{th:2} and Corollary \ref{co:1} indicate that in (\ref{eq:3.3}), the 2-D frequency information is coded into $\mathbf{u}_x(\mathbf{f}_x)$ and $\mathbf{u}_y(\mathbf{f}_y)$ in a decoupled manner. } 
Decoupling greatly reduces the overall computational complexity, which will be analyzed in detail in Section V.D. Indeed, the PSD matrix in  (\ref{eq:3.3}) is of size $(N+M)\times(N+M)$, which is markedly smaller than that of the vectorization approach in (\ref{eq:2.6}). In addition to its advantage in complexity, the decoupling in (\ref{eq:3.3}) retains the performance benefits of joint 2-D frequency estimation, with no loss of optimality as indicated by Theorem \ref{th:1}. {{Unlike the suboptimal decoupling strategy in Section \ref{sect:suboptimal}}, the two frequency dimensions are still coupled in the atomic decomposition in (\ref{eq:3.2}), so that both $\mathbf{u}_x(\mathbf{f}_x)$ and $\mathbf{u}_y(\mathbf{f}_y)$  are jointly retrieved from the data $\mathbf{X}$ in (\ref{eq:3.3}).
It is the new matrix-form atom set (\ref{eq:3.1}) that naturally results in the decoupled SDP for optimization, without invoking any relaxation or approximation.

\section{Decoupled ANM for 2-D Line Spectrum Estimation}
\label{sec:danm}

{Theorem \ref{th:2} and Corollary \ref{co:1} suggest that 2-D line spectrum estimation can be carried out in two steps. First, the SDP in (\ref{eq:3.3}) yields $\mathbf{u}_x^{\star}$ and $\mathbf{u}_y^{\star}$ and hence $\mathbf{T}(\mathbf{u}_x^\star)$ and $\mathbf{T}(\mathbf{u}_y^{\star})$. Second, mature techniques for Vandemonde decomposition of one-level Teoplitz matrices can be employed to recover  $\mathbf{f}_x$ and $\mathbf{f}_y$ separately, followed by frequency pairing. 
}   

\subsection{D-ANM Formulation}

In practice, we usually do not have $\mathbf{X}$ at hand, but observe it from a compressed and/or noisy measurement $\mathbf{Y}$ via (\ref{eq:2.7}). 
Given $\mathbf{Y}$,  the ANM formulation in (\ref{eq:an3}) can be adopted, which leads to a regularized de-noising formulation to recover $\mathbf{X}$:
\begin{equation}
	\label{eq:4.1.1}
	\min_{\hat{\mathbf{X}}} \left\{\lambda\|\hat{\mathbf{X}}\|_{\mathcal{A}_M}+\|\mathbf{Y}-\mathcal{C}(\hat{\mathbf{X}})\|^2 \right\}.
\end{equation}
Here $\|\hat{\mathbf{X}}\|_{\mathcal{A}_M}$ is the sparsity-enforcing term, $\|\mathbf{Y}-\mathcal{C}(\hat{\mathbf{X}})\|^2$ is the noise-controlling term, and $\lambda$ is a weighting parameter. 

Utilizing Theorem \ref{th:2}, under the assumptions of Theorem \ref{th:2} { (\ref{eq:4.1.1}) can be equivalently written in a decoupled SDP formulation as follows}:
\begin{equation}
	\label{eq:3.4}
		\begin{split}
		&\min_{\hat{\mathbf{X}}, \hat{\mathbf{u}}_x, \hat{\mathbf{u}}_y} \left\{\frac{\lambda}{2\sqrt{NM}}\bigg(\mathrm{trace}\big(\mathbf{T}(\hat{\mathbf{u}}_x)\big)+\mathrm{trace}\big(\mathbf{T}(\hat{\mathbf{u}}_y)\big)\bigg)\right. \\
		&+\|\mathbf{Y}-\mathcal{C}(\mathbf{\hat{X}})\|_\mathrm{F}^2 \bigg\}  ~~~~ \mathrm{s.t.}~ \left(\begin{array}{cc}
			\mathbf{T}(\hat{\mathbf{u}}_y) & \hat{\mathbf{X}}^\mathrm{H} \\
			\hat{\mathbf{X}} & \mathbf{T}(\hat{\mathbf{u}}_x)
		\end{array}\right)\succeq \mathbf{0}.
	\end{split}
\end{equation}
The SDP in (\ref{eq:3.4}) can be solved efficiently by popular convex optimization toolboxes. {We term (\ref{eq:3.4}) as the decoupled ANM (D-ANM) formulation, because it decouples the 2-D frequency information into $\mathbf{u}_x$ and $\mathbf{u}_y$ in calculating the atomic norm. For the single-snapshot case, (\ref{eq:3.4}) is key in constructing the well-structured Toeplitz matrices for ensuing frequency estimation.}

\subsection{Frequency Extraction}

Upon solving (\ref{eq:3.4}),  the optimal estimate $\hat{\mathbf{u}}_x$ leads to an $N\times N$ Toeplitz matrix $\mathbf{T}(\hat{\mathbf{u}}_x)$, which reveals $\hat{\mathbf{f}}_x$ via one-level Vandermonde decomposition as follows:
\begin{equation}
	\label{eq:Tx}
	\mathbf{T}(\hat{\mathbf{u}}_x)=\mathbf{A}_N(\hat{\mathbf{f}}_x) \mathbf{D}_x \mathbf{A}_N^{\mathrm{H}}(\hat{\mathbf{f}}_x), \quad \mathbf{D}_x \succeq \mathbf{0} \mathrm{~is~diagonal}. 
\end{equation}
Similarly,  $\hat{\mathbf{f}}_y$ is coded in the $M\times M$ matrix $\mathbf{T}(\hat{\mathbf{u}}_y)$.  There are many mature techniques for extracting  $\hat{\mathbf{f}}_x$ and $\hat{\mathbf{f}}_y$ from the corresponding Toeplitz matrices, such as subspace methods, matrix pencil \cite{yang2016vandermonde} and Prony's method \cite{yang2015gridless}. Solving such one-level Vandermonde decomposition is much simpler than the two-level Vandermonde decomposition needed for the vectorized ANM approach  \cite{chi2015compressive, yang2016vandermonde}.

\subsection{Frequency Pairing}

Like many 2-D line spectrum estimation methods, a pairing step is critical in order to identify the $L$ frequency pairs $(\hat{f}_{x, l}, \hat{f}_{y, l})$, $\forall l$. 
Next we develop a simple pairing technique, utilizing the fact that we have $\hat{\mathbf{X}}$ at hand after solving (\ref{eq:3.4}). 

Note the signal model in (\ref{eq:2.2}) that $\mathbf{X}=\mathbf{A}_N(\mathbf{f}_x)\mathbf{S}\mathbf{A}_M^\mathrm{H}(\mathbf{f}_y)$.
Given $\hat{\mathbf{f}}_x, \hat{\mathbf{f}}_y$ and $\hat{\mathbf{X}}$, we define ($(\cdot)^{\dagger}$ denotes pseudo-inverse)
\begin{equation}
	\label{eq:3.6}
	\hat{\mathbf{S}}=\mathbf{A}_N^\dagger(\hat{\mathbf{f}}_x)\hat{\mathbf{X}}\big(\mathbf{A}_M^\mathrm{H}(\hat{\mathbf{f}}_y)\big)^\dagger,
\end{equation}
which is a re-ordered version of $\mathbf{S}$ if the recovery is perfect. 
If all frequencies are non-overlapping in both dimensions, then $\mathbf{T}(\hat{\mathbf{u}}_x)$ and $\mathbf{T}(\hat{\mathbf{u}}_y)$ have the same rank,  $\mathbf{S}$ is a diagonal matrix, and the re-ordered $\hat{\mathbf{S}}$ has only up to one non-zero element in each row or column. if some frequency components are overlapped in one dimension while separated in the other dimension as indicated by the worst-case scenario of the separation condition in Theorem \ref{th:2}, then $\mathrm{rank}(\mathbf{T}(\hat{\mathbf{u}}_x))\neq\mathrm{rank}(\mathbf{T}(\hat{\mathbf{u}}_y))$, $\hat{\mathbf{S}}$ may have multiple non-zero elements in either its row or column, but not both. In both cases, $\hat{f}_{x, i}$ should be paired with $\hat{f}_{y, j}$ if $\hat{s}_{ij}\neq 0$, without raising ambiguity. 

In the presence of noise, the pairing criteria based on $\hat{\mathbf{S}}$ can be improved. Suppose that  $\mathrm{rank}(\mathbf{T}(\hat{\mathbf{u}}_x))\geq \mathrm{rank}(\mathbf{T}(\hat{\mathbf{u}}_y))$ without losing of generality, which suggests that there is no frequency overlapping at least along the $x$ dimension. {That is,  $L_x = L$ and $L_y\leq L$.} In this case, we note that $\mathbf{D}_x$ in (\ref{eq:Tx}) is diagonal and positive definite, which means that the matrix $\mathbf{P}:=|\mathbf{D}_x^{-1}\hat{\mathbf{S}}|$ shares the same structure as $\hat{\mathbf{S}}$, and all its non-zero elements are close to $1$.  Hence, $\mathbf{P}$ can be used for pairing as well, with added noise resilience. Specifically, if an element  $p_{ij}:=[\mathbf{P}]_{(i,j)}$ exceeds a prescribed threshold $\epsilon \in (0, 1)$, then it is declared as being non-zero and $(\hat{f}_{x, i}, \hat{f}_{y, j})$ are paired. Overall, the pairing rule is summarized in Algorithm \ref{alg:pair2}. \begin{algorithm}[h]
	\caption{Frequency pairing for D-ANM} 
	\label{alg:pair2}
	\begin{algorithmic}[1]
		\REQUIRE
		Retrieved frequencies without pairing, $\hat{\mathbf{f}}_x, \hat{\mathbf{f}}_y$; \\
		Vandermonde matrix, $\mathbf{T}(\hat{\mathbf{u}}_x)$; \\
		Recovered signal of interest, $\hat{\mathbf{X}}$; \\
		Threshold $\epsilon$;
		\ENSURE
		\STATE  Construct the manifold  matrix $\mathbf{A}_N(\hat{\mathbf{f}}_x)$ from $\hat{\mathbf{f}}_x$;
		\STATE  Construct the manifold matrix $\mathbf{A}_M(\hat{\mathbf{f}}_y)$ from $\hat{\mathbf{f}}_y$;
		\STATE Compute $\mathbf{D}_x$ in (\ref{eq:Tx});
		\STATE Compute $\hat{\mathbf{S}}$ in (\ref{eq:3.6});
		\STATE Compute $\mathbf{P}=|\mathbf{D}_x^{-1}\hat{\mathbf{S}}|$;
		\RETURN $(\hat{f}_{x, i}, \hat{f}_{y, j}), \forall p_{ij}\geq\epsilon$.
	\end{algorithmic}
\end{algorithm}


The performance of this simple pairing algorithm depends on the noise level and the threshold $\epsilon$. It is applicable even when some frequencies overlap on one dimension, as long as the separation condition in Theorem \ref{th:2} holds. Other pairing techniques can be employed as well \cite{icassp}, and the choice depends on the tradeoff between complexity and accuracy in the presence of closely-spaced sources and weak signals.



\section{Analysis}
\label{sec:dis}

This section analyzes several properties of the proposed D-ANM method.




\subsection{Compression}

Consider the use of linear compression in collecting the measurement $\mathbf{Y}$. 
Without loss of generality, we rewrite the linear compressive operator $\mathcal{C}(\cdot)$ in (\ref{eq:2.7}) in the form 
\begin{equation}
	\label{eq:6.1}
		\mathcal{C}(\mathbf{X})=\mathbf{C}_x\mathbf{X}\mathbf{C}_y^\mathrm{H}
		=\sum_{l=1}^{L} s_l \left(\mathbf{C}_x\mathbf{a}_N(\mathbf{f}_x)\right)\left(\mathbf{C}_y\mathbf{a}_M(\mathbf{f}_y)\right)^\mathrm{H},
\end{equation} 
where $\mathbf{C}_x\in\mathbb{C}^{K_x\times N}$ and $\mathbf{C}_y\in\mathbb{C}^{K_y\times M}$ are the compression matrices in the $x$ and $y$ dimensions respectively, and $K_x (\leq N)$ and $K_y (\leq M)$ are the corresponding numbers of sample points along these two dimensions. Hence the compression ratio is $\rho = (K_xK_y)/(NM)$.

In the noise-free case, the compression performance of D-ANM is given in the following theorem.
\begin{theorem}
	\label{th:3} 
	Suppose that the following conditions hold: 
	\begin{itemize}
		\item $\mathbf{C}_x$ and $\mathbf{C}_y$ are both random matrices and $N, M\geq  512$;
		\item $\mathbb{E}(\mathbf{C}_x^\mathrm{H}\mathbf{C}_x)=\mathbf{I}_N, \mathbb{E}(\mathbf{C}_y^\mathrm{H}\mathbf{C}_y)=\mathbf{I}_M$ are identity matrices;
		\item $\mathbb{E}(\mathbf{c}_{x, k_x}\mathbf{c}_{x, k_x}^\mathbf{H})=\frac{1}{N}\mathbf{I}_N, \forall k_x \in [1, K_x]$;
		\item $\mathbb{E}(\mathbf{c}_{y, k_y}\mathbf{c}_{y, k_y}^\mathbf{H})=\frac{1}{M}\mathbf{I}_M, \forall k_y \in [1, K_y]$;
		\item $\sup_{\mathbf{A}(\mathbf{f})\in\mathcal{A}} \left\|\left\langle \mathbf{A}(\mathbf{f}),  \mathbf{c}_{x, k_x}\mathbf{c}_{y, k_y}^\mathbf{H}\right\rangle\right\|^2\leq \mu\frac{NM}{K_x K_y}$, {$\forall k_x, k_y$};
		\item $\mathbf{f}$ satisfies the sufficient separation condition (\ref{eq:2.5})
	\end{itemize}
	where $\mathbf{c}_{x, k_x}$ is the $k_x$-th column of $\mathbf{C}_x$, $\mathbf{c}_{y, k_y}$ is the $k_y$-th column of $\mathbf{C}_y$, and $\mu\geq 1$ is a constant. Then, the ANM formula in (\ref{eq:3.4}) reveals the true frequencies with at least probability  $1-\delta$ as long as
	\begin{equation}
		K_x K_y \geq C \mu L\log\left(\frac{N+M}{\delta}\right),
	\end{equation}
	where $C$ is a constant.
\end{theorem}

The proof is inspired by \cite{heckel2016generalized} [Theorem 1], with proper adjustments for extension to the {2-D} case.  Remark that this result also works for the 1-D case with $M=1$. 

Note that conditions of this theorem require the entries of compression matrices to be statistically orthogonal and uncorrelated, and {have bounded values.}  Some widely used random compression matrices such as Gaussian matrices do not satisfy this conditions because they are unbounded, even though they usually work well in practice. Meanwhile, Bernoulli and sparse sensing matrices such as nested samplers \cite{7113829} naturally satisfy the conditions. 

\subsection{Frequency Resolution}

Here resolution refers the required frequency separation between neighboring sinusoidal components. In ANM-based methods, this is described by the sufficient separation condition, as in (\ref{eq:2.5}).

In vectorized ANM, the sufficient separation condition is looser than that in D-ANM. For example, suppose $N=M$, the condition of vectorized ANM is
\begin{displaymath}
	\min_{i\neq j} \max \left\{|f_{x, i}-f_{x, j}|, |f_{y, i}-f_{y, j}|\right\}\geq \frac{1}{\lfloor(N-1)/4\rfloor}.
\end{displaymath}
If any pair of two sources are separated enough in at least one dimension, vectorized ANM allows $\Delta_{\min, x}$ and $\Delta_{\min, y}$ to be both less than $\frac{1}{\lfloor(N-1)/4\rfloor}$. This condition is weaker than that of D-ANM in (\ref{eq:2.5}), making vectorized ANM applicable in broader scenarios in theory. However, the described sufficient separation conditions are rather conservative and tend to be loose bounds, whereas in most realistic  applications, both ANM methods do not exhibit  evident difference in terms of the frequency resolution.



One may also notice that (\ref{eq:2.5}) is much more conservative than the resolution condition of subspace methods \cite{pesavento2004multidimensional, liu2006eigenvector}, which is determined by the array aperture. It is worth noting that 
\begin{itemize}
	\item In subspace methods, the resolution determined by Rayleigh aperture is usually strict and cannot be violated.
	\item In ANM, (\ref{eq:2.5}) is only a sufficient condition, and usually very conservative in practice. That is, even if (\ref{eq:2.5}) is not satisfied, one may still have a chance to retrieve all frequencies with high probability. We will show this in the simulation section.
\end{itemize} 


\subsection{Number of Sources (Identifiability)}

Similar to the CS approach, the number of sinusoidal sources $L$, also termed as \emph{signal sparsity} in the frequency domain, impacts the recovery performance of ANM. It is easy to see that in order to guarantee a unique Vandermonde decomposition, the size of a Toeplitz matrix should be at least larger than its rank. Hence, the maximum number of recoverable sources is limited by
\begin{equation}
	\label{eq:5.1}
	L\leq\min\{N, M\}.
\end{equation}
This is the same to that of the vectorized approach, as stated in \cite{chi2015compressive} [Proposition 2]. 

In contrast, smoothing-based subspace methods \cite{liu2006eigenvector, liu2006statistical} guarantee identifiability for at least
\begin{displaymath}
	L\leq 0.34NM.
\end{displaymath}
Hence, smoothing-based subspace methods are advantageous in terms of the identifiability property. Such an advantage is more evident when the dimension grows larger \cite{liu2006statistical}. On the other hand,  the identifiability of smoothing-based subspace methods for the single-snapshot case is achieved by reducing the effective aperture size, which results in reduced resolution. The ANM methods, on the other hand, retain the benefits of full aperture size.

It is worth noting that the limitation of $L\leq\min\{N, M\}$ in ANM is induced by the SDP and Vandermonde decomposition step. It is not imposed by performing atomic norm minimization. This suggests that we might be able to bypass this limitation, if we could find an alternative way to calculate the atomic norm and retrieve the frequencies, other than SDP and Vandermonde decomposition.


\subsection{Computational Complexity}

Most SDP solvers are programed based on the interior point method,
for which the complexity is studied in \cite{krishnan2005interior}. Specifically, the SDP solver needs approximately $\mathcal{O}(P^3+P^2)=\mathcal{O}(P^3)$ steps for each iteration, and at most $\mathcal{O}(\sqrt{P}\log(1/\epsilon))$ iterations, where $P$ is the size of the PSD  matrix in the constraint and $\epsilon$ is the desired recovery precision. Hence, the overall time complexity becomes $\mathcal{O}(P^{3.5}\log(1/\epsilon))$.

After SDP, Vandermonde decomposition is needed to retrieve all frequencies. For vectorized ANM, a 2-level Toeplitz decomposition \cite{yang2016vandermonde} is required which has computational complexity  $\mathcal{O} (P^2 L)$, where $P$ is the size of 2-level Toeplitz matrix and $L$ is its rank. For our proposed D-ANM method, only two separate 1-level Toeplitz matrix Vandermonde decompositions are required at complexity $\mathcal{O}(P^2)$ \cite{backstrom2013vandermonde}.

For vectorized ANM \cite{chi2015compressive}, the constraint size is $P=NM+1$, and for  D-ANM, the constraint size becomes $P=N+M$. The complexity comparison of these two  methods  is listed in Table \ref{tb:1}. It is evident that  D-ANM remarkably reduces the complexity by an order of $N^{3.5}$ (if $N$ and $M$ are on the same order), which is significant for large values of $N$ and $M$.
\begin{table}[h]
	\centering
	\caption{Complexity}
	\begin{tabular}{|p{100pt}||p{100pt}|}
		\hline
		~ & Complexity \\
		\hline
		Vectorized ANM & $\mathcal{O}(N^{3.5}M^{3.5}\log(1/\epsilon))$ \\
		2-Level V-decomposition & $\mathcal{O}(N^2 M^2 K)$ \\
		\hline
		D-ANM & $\mathcal{O}((N+M)^{3.5}\log(1/\epsilon))$ \\
		1-Level V-decomposition & $\mathcal{O}(N^2+M^2)$ \\
		\hline
	\end{tabular}
	\label{tb:1}
\end{table}

\section{Numerical Results}
\label{sec:sim}

In this section, we use simulations to validate our proposed D-ANM, compared to  the existing vectorized ANM method. We apply the matrix pencil method \cite{yang2016vandermonde} for both 1-D and 2-D Vandermonde decomposition. If not specifically stated, the default simulation settings are listed in Table \ref{table:1}. The algorithms are implemented using CVX toolbox \cite{grant2008cvx}. 

\begin{table}[h!]
	\centering
	\caption{The simulation settings}
	\centering
	\label{table:1}
	\begin{tabular}{|l||l|}
		\hline
		Parameter & Value \\
		\hline
		$N$ & 16 \\
		$M$ & 16 \\
		$L$ & 4 \\
		$f_x$ & Random, uniformly generated \\
		$f_y$ & Random, uniformly generated \\
		$s_l$ & Random, Gaussian generated\\
		SNR & 20dB \\
		Compression & No compression, unless stated \\
		\hline
	\end{tabular}
\end{table}

\subsection{Run Time}

We firstly compare the run time of our proposed method versus the vectorized ANM. Simulations are performed on a square array with $M=N$ varying from 8 to 22. As shown in Figure \ref{fig:2}, Our method exhibits a huge advantage in computational efficiency for large-scale arrays. When $M=N=22$, the running time of the vectorized ANM is 733.1842s, while that of the D-ANM is only 1.4997s.

\begin{figure}[h]
	\centering
	\includegraphics[width=.9\columnwidth]{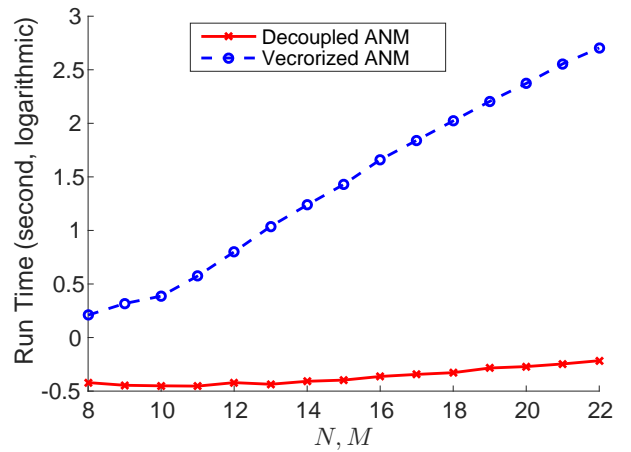}	
	\caption{Computing complexity: run time versus $N$ ($N=M$). Time scale: $\log_{10}$second}
	\label{fig:2}
\end{figure}

\subsection{Performance Comparison}

We now compare the performance of our proposed D-ANM with other benchmark methods, including CS and vectorized ANM. The CS method was solved on a $16\times 16$ grid.

In the simulation, we test the mean square error (MSE) performance of recovered $\mathbf{f}$ versus SNR for each method, with uncompressed data. As Figure \ref{fig:7} shows, the MSE performance of the proposed D-ANM is quite close to that of the vectorized ANM, and both of them approach the Cramer-Rao bound (CRB) \cite{liu2007multidimensional, stoica1989music}  when SNR is high. However, the CS method performs much worse because its precision is limited by the grid size regardless of the SNR range.

\begin{figure}[t]
	\centering
		\includegraphics[width=.9\columnwidth]{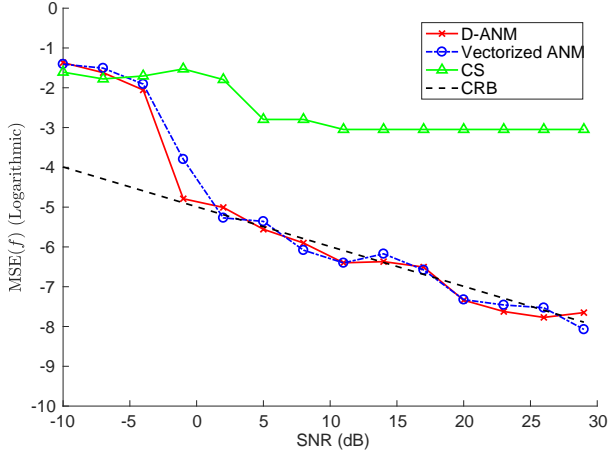}
	\caption{Comparison of MSE performance versus SNR ($N=M=16$).}
	\label{fig:7}
\end{figure}

\subsection{Frequency Separation}

In order to observe how sensitive the performance of ANM is with respect to the frequency separation $\Delta$, we test in Figure \ref{fig:5} the successful recovery rate versus the frequency separation of two sources, that is, $L=2$. We define a successful recovery for the recovered frequency when $\mathrm{MSE}\leq 10^{-6}$ in the noise-free case. Given $N=M=16$,  the theoretical minimal separation in (\ref{eq:2.5}) is $\Delta_{\min}\geq \frac{1}{\lfloor(N-1)/4\rfloor}=0.33$. However, Fig. \ref{fig:5} shows that the ANM performs well even when $\Delta$ is well below $0.33$. Even when $\Delta_{\min}$ is as small as 0.1, There is still a high probability of successful recovery. The D-ANM fails at $\Delta_{\min}<0.05$ in this experiment, at which point the successful recovery rate rapidly drops to 0. 

Indeed, the condition (\ref{eq:2.5}) is a sufficient but not necessary condition. This is a useful guideline for ANM practice, so that we do not need to be overly concerned with the frequency separation even if (\ref{eq:2.5}) is not satisfied.
\begin{figure}[h]
	\centering
	\includegraphics[width=.9\columnwidth]{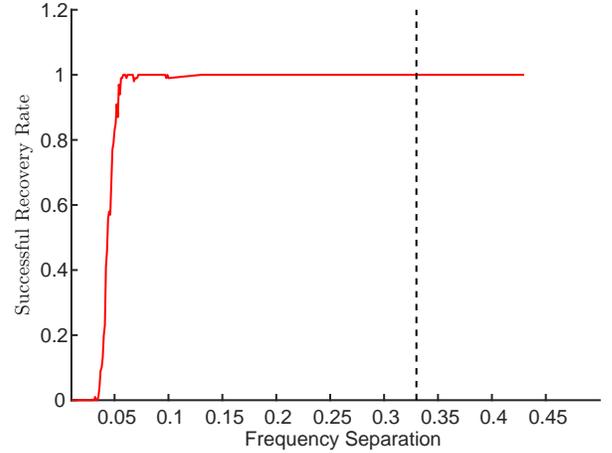}
	\caption{Successful recovery rate of D-ANM versus the minimum frequency separation for $L=2$.}
	\label{fig:5}
\end{figure}

\subsection{Compression}
\label{sect:compression}

Under data compression, the performance of D-ANM is tested against the benchmark CS and vectorized ANM methods. As explained in (\ref{eq:6.1}), linear compression matrices in the form of random Bernoulli matrices are applied on both sides of the signal $\mathbf{X}$, with $K_x = K_y$ varying from 2 to 16. The corresponding compression ratio $\rho = K_xK_y/(NM)$ varies from 1.5\% to 100\%. The resulting MSEs are depicted in Figure \ref{fig:3}. Whereas the performance of CS is limited by the gridding leakage effect, both ANM methods perform very close and show  improved performance as the compression ratio increases. A sharp performance change at $\rho=20\%$ indicated the phase transit point, which will be discussed next via phase transition diagrams.  

\begin{figure}[h]
	\centering
	\includegraphics[width=.9\columnwidth]{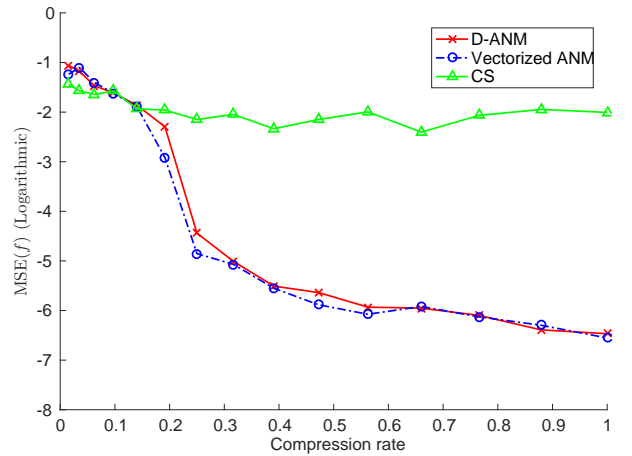}
	\caption{Comparison of MSE performance versus compression ratio ($N=M=16$).}
	\label{fig:3}
\end{figure}

\subsection{Number of Sources (Sparsity)}

We now test the performance of the D-ANM method as the number of sources varies. In order to separate the effect of sparsity from that of  frequency separation, the minimum frequency separation is fixed at $\Delta_{\min}=0.05$ in the test, regardless of $L$. 

Figure \ref{fig:4} depicts the recovery MSE when $L$ varies from 2 to 7, with the CRB as reference. It is evident that the MSE is quite close to CRB when $L$ is small, which verifies the effectiveness of the D-ANM method. When $L>4$, the MSE jumps up significantly, which indicates that the D-ANM fails to work in this region of $L$. In another words, $L=4$ is the transition point between the two phases ``success'' and ``failure'' in line spectrum estimation using the D-ANM. Since $N=M=16$, theoretically the D-ANM should be able to identify up to $15$ sources according to (\ref{eq:5.1}); on the other hand, the frequency separation condition in (\ref{eq:2.5}) is subject to an increasing probability of being violated as $L$ exceeds $4$. Apparently, the location of the phase transition point is greatly affected by the problem sizes $N$ and $M$ under the frequency separation condition. 

The phase transition diagram in Figure \ref{fig:8}  depicts  the empirical probability that the D-ANM successfully identifies and estimates the 2-D DOA of $L$ randomly placed sources, as $N=M$ increases. The shaded area in the lower left corner indicate successful recovery, whereas the white area in the upper right corner indicates failure. A phase transition border between these two areas can be observed, which consists of all the phase transit points. As $N$ and $M$ increases, the phase transition point of $L$ also increases.

In summary, sparsity critically influences the ANM family for line spectrum estimation. As a structure-based method, the ANM  enjoys the benefits of applicability with only one-snapshot measurement and robustness to source correlation, at the expense of a limited number of sources due to the sparsity constraint. In contrast, at $N=M=16$, a subspace method can separate up to $L=15$ sources, assuming the availability of a large number of snapshots.

\begin{figure}[h]
	\centering
	\includegraphics[width=.9\columnwidth]{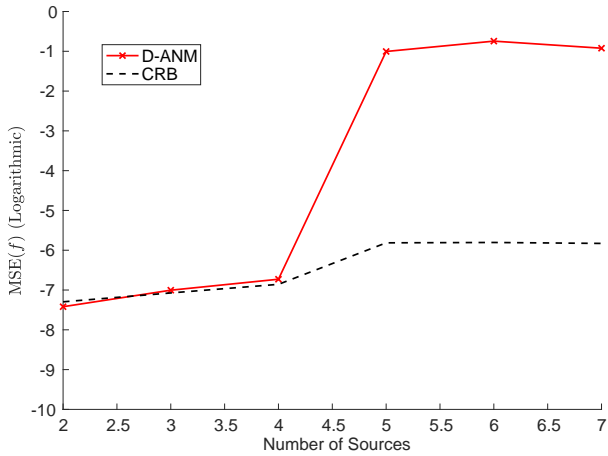}
	\caption{MSE performance of D-ANM versus the number of sources $L$.}
	\label{fig:4}
\end{figure}

\begin{figure}[h]
	\centering
	\includegraphics[width=.9\columnwidth]{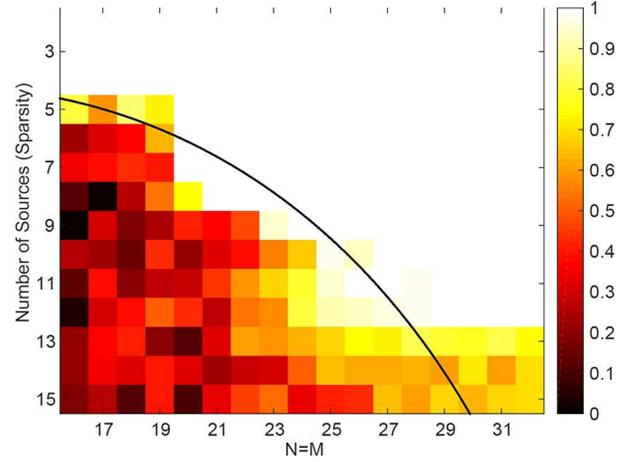}
	\caption{Phase transition diagram of  $L$ and  $N(=M)$.}
	\label{fig:8}
\end{figure}

\subsection{Phase Transition between Compression and Sparsity}

For the ANM, the phase transition phenomenon also emerges between the compression and sparsity. Linear compression is done on both dimensions, as in Section \ref{sect:compression}. Figure \ref{fig:6} depicts the probability of successful recovery as the number of sources $L$ and the compression ratio $\rho$ vary. 
Apparently, the probability of successful recovery is critically affected by the sparsity level, and appears to be less sensitive to the compression ratio. 
\begin{figure}[h]
		\centering
		\includegraphics[width=.9\columnwidth]{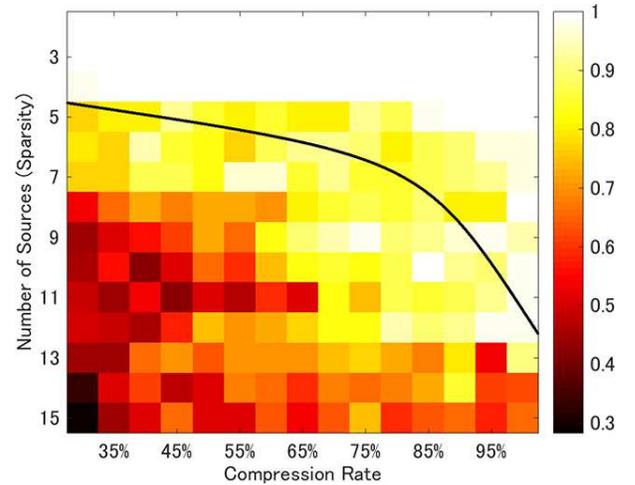}
		\caption{Phase transition diagram of compression and sparsity $L$.}
		\label{fig:6}	
\end{figure}

\section{Conclusion}
\label{sec:con}

This paper develops a computationally efficient decoupled-ANM approach for 2-D line spectrum estimation. The idea is to use rank-one matrix-form atoms of size $N\times M$ to replace the conventional vectorized atom set. With rigorous proofs, we have shown that if the sinusoids are sufficiently separated in the frequency domain, one can successfully recover the frequencies via a decoupled SDP formulation. This method also yields the true signal with high probability in the compressive case.
Compared with the conventional vectorization-based ANM approach, the proposed D-ANM dramatically reduces the problem scale from $NM\times NM$ to $(N+M)\times(N+M)$, which results in reduction of complexity on the order of $N^{3.5}$ for $N=M$, and retains the benefits of conventional ANM with little performance degradation in the noisy case. This makes the D-ANM practical in applications with a large problem scale, such as massive MIMO, radar signal processing and radio astronomy.


\appendices

\section{Proof of Theorem \ref{th:1}}
\label{sec:app1}


The following steps are elucidated to prove Theorem \ref{th:1}:
\begin{enumerate}
	\item We show that, if there exists an appropriate polynomial (say dual polynomial) that satisfies some given conditions, then Theorem \ref{th:1} holds.
	\item We show the existence of dual polynomial by constructing it.
\end{enumerate}
These steps follow the similar approach in the 1-D case \cite{tang2013compressed}, and that in the 2-D vectorized case \cite{chi2015compressive}.

\subsection{Dual Certificate}
\label{sec:app1.1}

%


$\forall\mathbf{Q}\in\mathbb{C}^{N\times M}$, define the dual norm of $\|\cdot\|_{\mathcal{A}_M}$ as

\begin{equation}
\label{eq:a3}
\|\mathbf{Q}\|_{\mathcal{A}_M}^\ast=\sup_{\|\mathbf{X}\|_{\mathcal{A}_M}\leq 1} \langle\mathbf{Q}, \mathbf{X}\rangle_\mathcal{R}=\sup_{{\mathcal{A}_M}\in\mathcal{A}} \Re\left\{ \mathrm{trace}( \mathbf{A}^\mathbf{H} \mathbf{Q})\right\}.
\end{equation}
where, $\langle\cdot, \cdot\rangle$ denotes the Frobenius inner product and
$\langle\cdot, \cdot\rangle_\mathcal{R}=\Re\left\{\langle\cdot, \cdot\rangle\right\}$.

Follow the standard Lagrangian procedure as in \cite{Chandrasekaran2012}, we can find the dual problem of D-ANM as
\begin{equation}
\label{eq:a1}
\max_{\mathbf{Q}} \langle \mathbf{Q}, \mathbf{X}\rangle_\mathcal{R},\mathrm{~~~~~s.t.~}\|\mathbf{Q}\|_{\mathcal{A}_M}^\ast\leq 1.
\end{equation}
Since the original convex problem is unconstrained, strong duality naturally holds. This motivates the concept of dual polynomial by studying the constraint of (\ref{eq:a1}).

A lemma is introduced to show the relationship between the uniqueness and dual polynomial. This lemma generalizes the results in \cite{candes2014towards, tang2013compressed} to the matrix form.

\begin{lemma}
	\label{lm:2}
	Consider a matrix $\mathbf{X}\in\mathbb{C}^{N\times M}$ in the form
	\begin{equation}
	\label{eq:u1}
	\mathbf{X}=\sum_{l=1}^L s_l \mathbf{A}(\mathbf{f}_l),\mathbf{~~~}\mathbf{A}(\mathbf{f}_l)\in\mathcal{A}_M,
	\end{equation}
	and let the set $\Omega=\{f_l, l=1, \dots, L\}$ collect all the frequency supports of $\mathbf{X}$.
	
	If there exists a dual polynomial in the form of
	\begin{equation}
	\label{eq:A.1}
	Q(\mathbf{f})=\langle \mathbf{Q}, \mathbf{A}(\mathbf{f})\rangle=\mathrm{trace}(\mathbf{A}^\mathrm{H}(\mathbf{f})\mathbf{Q}),
	\end{equation}
	which is amenable for some $\mathbf{Q}\in\mathbb{C}^{N\times M}$ that satisfies the conditions
	\begin{equation}
	\label{eq:c1}
	Q(\mathbf{f}_l)=\mathrm{sign}(s_l), \mathrm{~~~}\forall \mathbf{f}_l\in\Omega,
	\end{equation}
	\begin{equation}
	\label{eq:c2}
	|Q(\mathbf{f})|<1, \mathrm{~~~} \forall \mathbf{f}\notin\Omega,
	\end{equation}
	then it is guaranteed that (\ref{eq:u1}) is the unique optimal atomic decomposition of $\mathbf{X}$, that is,
	\begin{equation}
	\|\mathbf{X}\|_{\mathcal{A}_M}=\sum_{l=1}^L |s_l|.
	\end{equation}
\end{lemma}

{
	Proof: From H\"{o}lder's inequality, we have
	\begin{equation}
	\label{eq:lm1}
	\langle\mathbf{Q}, \mathbf{X}\rangle\leq\|\mathbf{Q}\|_{\mathcal{A}_M}^\ast \|\mathbf{X}\|_{\mathcal{A}_M}.
	\end{equation}
	Since $|Q(\mathbf{f})|\leq 1$ holds for $\forall \mathbf{f}\in[0, 1]\times[0, 1]$,
	\begin{equation}
	\label{eq:lm2}
	\begin{split}
	\|\mathbf{Q}\|_{\mathcal{A}_M}^\ast&=\sup_{\mathbf{A}(\mathbf{f})\in{\mathcal{A}_M}} \langle\mathbf{Q}, \mathbf{A}(\mathbf{f})\rangle_\mathcal{R} \\
	&=\sup_{\mathbf{A}(\mathbf{f})\in{\mathcal{A}_M}} \Re\{Q(\mathbf{f})\} \leq 1.
	\end{split}
	\end{equation}
	It implies from (\ref{eq:lm1}) and (\ref{eq:lm2}) that
	\begin{equation}
	\label{eq:lm3}
	\langle\mathbf{Q}, \mathbf{X}\rangle\leq \|\mathbf{X}\|_{\mathcal{A}_M}.
	\end{equation}
	
	On the other hand,
	\begin{equation}
	\label{eq:lm4}
	\begin{split}
	\langle \mathbf{Q}, \mathbf{X}\rangle&=\left\langle\mathbf{Q}, \sum_l s_l \mathbf{A}(\mathbf{f}_l) \right\rangle \\
	&=\mathrm{trace} \left( \left(\sum_l s_l^\ast \mathbf{A}^\mathrm{H}(\mathbf{f}_l)\right) \mathbf{Q} \right) \\
	&=\sum_l s_l^\ast \mathrm{trace} \left(  \mathbf{A}^\mathrm{H}(\mathbf{f}_l) \mathbf{Q} \right) \\
	&=\sum_l s_l^\ast \left( Q(\mathbf{f}_l) \right) =\sum_l s_l^\ast \mathrm{sign}(s_l) \\
	&=\sum_l |s_l| \geq \|\mathbf{X}\|_{\mathcal{A}_M},
	\end{split}
	\end{equation}
	which holds because of the definition of atomic norm.
	
	From (\ref{eq:lm3}) and (\ref{eq:lm4}), we conclude that (\ref{eq:u1}) is an optimal decomposition of $\mathbf{X}$, and 
	\begin{equation}
	\|\mathbf{X}\|_{\mathcal{A}_M}=\sum_l |s_l|. 
	\end{equation}
	
	Next, we will show the uniqueness of the above atomic decomposition.
	
	Suppose $\mathbf{X}$ has another atomic decomposition parameterized by $\{\hat{s}_k, \hat{\mathbf{f}}_k\}_{k=1}^K$, yielding
	\begin{equation}
	\mathbf{X}=\sum_k \hat{s}_k \hat{\mathbf{A}}(\mathbf{f}_k),
	\quad
	\|\mathbf{X}\|_{\mathcal{A}_M}=\sum_k |\hat{s}_k|.
	\end{equation}
	Then
	\begin{equation}
	\begin{split}
	\langle\mathbf{Q}, \mathbf{X}\rangle&=\left\langle\mathbf{Q}, \sum_k \hat{s}_k \hat{\mathbf{A}}(\mathbf{f}_k),  \right\rangle \\
	&=\sum_{\mathbf{f}_l\in\Omega} \hat{s}_l^\ast \left( Q(\mathbf{f}_l) \right)+\sum_{\mathbf{f}_k\notin\Omega} \hat{s}_k^\ast \left( Q(\mathbf{f}_k) \right) \\
	&< \sum_{\mathbf{f}_l\in\Omega} |s_l|+ \sum_{\mathbf{f}_k\notin\Omega} |\hat{s}_k| \\
	&=\|\mathbf{X}\|_{\mathcal{A}_M},
	\end{split}
	\end{equation}
	which causes contradiction with (\ref{eq:lm4}). Hence, the decomposition (\ref{eq:u1}) is unique. \QED
}

This lemma shows that Theorem \ref{th:1} holds if and only if we can construct a dual polynomial which satisfies (\ref{eq:c1}) and (\ref{eq:c2}). Next, our goal is to find one specific dual polynomial satisfying such conditions. 

\subsection{Construction of Dual Polynomial}
\label{sec:app1.2}


To show the existence, we start with an alternative model of $\mathbf{X}$ on a shifted frequency coordinate, that is, $\mathbf{f}\in[-\frac{1}{2}, \frac{1}{2}]^2$. This allows us to draw relevant results from the literatures \cite{chi2015compressive, candes2014towards}.

Specifically, let
\begin{equation}
\label{eq:p1}
\mathbf{X}=\sum_l \tilde{s}_l \tilde{\mathbf{A}}(\mathbf{f}_l), \mathrm{~~~~}\tilde{\mathbf{A}}(\mathbf{f})\in\tilde{\mathcal{A}}_{M},
\end{equation}
where the atoms are defined as
\begin{equation}
\tilde{\mathbf{A}}(\mathbf{f})=\tilde{\mathbf{A}}_{\tilde{N}}(f_x)\tilde{\mathbf{A}}_{\tilde{M}}^\mathrm{H}(f_y)\in\mathbb{C}^{(2\tilde{M}+1)(2\tilde{N}+1)\times 1},
\end{equation}
which columns as
\begin{displaymath}
\tilde{\mathbf{a}}_{\tilde{N}}(f)=\left(\exp(j\pi(-2\tilde{N})f), \dots, 1, \dots \exp(j\pi(2\tilde{N})f) \right)^\mathrm{T},
\end{displaymath}
\begin{displaymath}
\tilde{\mathbf{a}}_{\tilde{M}}(f)=\left(\exp(j\pi(-2\tilde{M})f), \dots, 1, \dots \exp(j\pi(2\tilde{M})f) \right)^\mathrm{T}.
\end{displaymath}

The dual polynomial is interpolated using Fej\'{e}r's Kernel. A 1-D Fej\'{e}r's kernel is defined as
\begin{equation}
\mathcal{K}_{\tilde{N}}(f)=\left(\frac{\sin(\pi\tilde{N}f)}{\tilde{N}\sin{\pi f}}\right)^4=\frac{1}{\tilde{N}}\sum_{n=-2\tilde{N}}^{2\tilde{N}}g_{\tilde{N}}(n)e^{-j2\pi f n},
\end{equation}
where $f\in[0, 1]$ and
\begin{equation}
g_{\tilde{N}} (n)=\frac{1}{\tilde{N}}\sum_{k=\max(n-\tilde{N}, -\tilde{N})}^{\min(n+\tilde{N}, \tilde{N})}\left(1-\left|\frac{k}{\tilde{N}}\right|\right)\left(1-\left|\frac{n-k}{\tilde{N}}\right|\right).
\end{equation}

In the 2-D case, the complex 2-D Fej\'{e}r's kernel is defined as
\begin{equation}
\label{eq:f1}
\begin{split}
\mathcal{K}(\mathbf{f})&=\mathcal{K}_{\tilde{N}}(f_x)\mathcal{K}_{\tilde{M}}^\ast(f_y) \\
&= \frac{1}{\tilde{N}\tilde{M}}\sum_{n}\sum_{m}g_{\tilde{N}}(n)g_{\tilde{M}}^\ast(m)e^{-j2\pi f_x n} e^{j2\pi f_y m}.
\end{split}
\end{equation}

Denote the partial derivative of $\mathcal{K}(\mathbf{f})$ as
\begin{equation}
\mathcal{K}^{(i, j)}(\mathbf{f})=\frac{\partial^i \partial^j \mathcal{K}(\mathbf{f})}{\partial f_x^i f_y^j}.
\end{equation}


The dual polynomial (\ref{eq:a3}) is constructed using an interpolation of 2-D Fej\'{e}r's kernel as
\begin{equation}
\label{eq:a2}
Q(\mathbf{f})=\sum_{l=1}^{L}\alpha_l \mathcal{K}(\mathbf{f}-\mathbf{f}_l) + \sum_{l=1}^{L}\beta_{1l} \mathcal{K}^{(1, 0)}(\mathbf{f}-\mathbf{f}_l) + \sum_{l=1}^{L}\beta_{2l} \mathcal{K}^{(0, 1)}(\mathbf{f}-\mathbf{f}_l),
\end{equation}
where $\alpha_l, \beta_{1l}, \beta_{2l}$ are interpolation coefficients. 
Similar to \cite{chi2015compressive} [Appendix B], these coefficients can be determined by solve a linear equation
\begin{equation}
\mathbf{E}\left(\begin{array}{c}
\boldsymbol{\alpha} \\
\frac{1}{K}\boldsymbol{\beta}_1 \\
\frac{1}{K}\boldsymbol{\beta}_2 \\
\end{array}\right)=\left(\begin{array}{c}
\mathrm{sign}(\mathbf{s}_l) \\
\mathbf{0} \\
\mathbf{0}
\end{array}\right),
\end{equation}
where
\begin{equation}
\mathbf{E}=\left(\begin{array}{ccc}
\mathbf{E}_{0, 0} & K\mathbf{E}_{1, 0} & K\mathbf{E}_{0, 1} \\
-K\mathbf{E}_{1, 0} & -K^2\mathbf{E}_{2, 0} & -K^2\mathbf{E}_{1, 1} \\
-K\mathbf{E}_{0, 1} & -K^2\mathbf{E}_{1, 1} & -K^2\mathbf{E}_{0, 2} 
\end{array}\right),
\end{equation}
\begin{equation}
K=\sqrt[4]{\mathcal{K}_N^{\prime\prime}(0) \mathcal{K}_M^{\prime\prime}(0)},
\end{equation}
and the $(k, l)$-th entry of $\mathbf{E}_{i, j}$ is
\begin{equation}
\bigg(\mathbf{E}_{i, j}\bigg)_{k, l}=\mathcal{K}^{(i, j)}(\mathbf{f}_k-\mathbf{f}_l).
\end{equation}

It has been shown in \cite{chi2015compressive} [Appendix C] that under the conditions of Theorem \ref{th:1}, the matrix $\mathbf{E}$ is invertible and has
\begin{equation}
\|\mathbf{I}-\mathbf{E}\|\leq 0.1982,
\end{equation}
where $\|\cdot\|$ denotes the matrix operator norm. Hence, the interpolation coefficients $\boldsymbol{\alpha}, \boldsymbol{\beta}_1, \boldsymbol{\beta}_2$ can be uniquely determined. 

Further, $\mathbf{E}$ can be expressed as
\begin{equation}
\mathbf{E}=\frac{1}{\tilde{N}\tilde{M}}\sum_{n=-2\tilde{N}}^{2\tilde{N}} \sum_{m=-2\tilde{M}}^{2\tilde{M}} g_{\tilde{N}}^\ast(n)  g_{\tilde{M}}(m) \mathbf{e}_{nm} \mathbf{e}_{nm}^\mathrm{H},
\end{equation}
where
\begin{equation}
\mathbf{e}_{nm}=\left(\begin{array}{c}
1 \\
j2\pi K n \\
j2\pi K m
\end{array}\right) \otimes \left(\begin{array}{c}
\exp(-j2\pi f_{x, 1} n )\exp(-j2\pi f_{y, 1} m) \\
\vdots \\
\exp(-j2\pi f_{x, L} n )\exp(-j2\pi f_{y, L} m)
\end{array}\right)
\end{equation}

\subsection{Shifting}
\label{sec:app1.3}

Finally, we have to shift the observing points from $\{-2\tilde{N}, \dots, 2\tilde{N}\}$, $\{-2\tilde{M}, \dots, 2\tilde{M}\}$ to  $\{0, \dots, 2N-1\}$, $\{0, \dots, 2M-1\}$, which follows similar steps as \cite{tang2013compressed} [Appendix A].

\section{Proof of Theorem \ref{th:2}}
\label{sec:app2}

{


Consider (\ref{eq:3.3}) in Theorem \ref{th:2}. Given the square Toeplitz matrices $\mathbf{T}(\mathbf{u}_x)$ and $\mathbf{T}(\mathbf{u}_y)$ parameterized by vectors $\mathbf{u}_x\in\mathbb{C}^N$ and $\mathbf{u}_y\in\mathbb{C}^M$ respectively, we define
\begin{displaymath}
	g(\mathbf{u}_x, \mathbf{u}_y) = \frac{1}{2\sqrt{NM}}\bigg(\!\mathrm{trace}\big(\mathbf{T}(\mathbf{u}_x)\big)\!+\!\mathrm{trace}\big(\mathbf{T}(\mathbf{u}_y)\big)\!\bigg).
\end{displaymath} 
For a given $\mathbf{X}\in\mathbb{C}^{N\times M}$, we denote a feasible set of $(\mathbf{u}_x, \mathbf{u}_y)$ as 
\begin{displaymath}
	S_{\mathbf{X}}^+(\mathbf{u}_x, \mathbf{u}_y) = \left\{(\mathbf{u}_x, \mathbf{u}_y) \bigg| \left(\begin{array}{cc}
	\mathbf{T}(\mathbf{u}_y) & \mathbf{X}^\mathrm{H} \\
	\mathbf{X} & \mathbf{T}(\mathbf{u}_x)
	\end{array}\right)\succeq \mathbf{0}
	\right\}.
\end{displaymath}
			
Let $\mathrm{SDP}(\mathbf{X})$ denote the optimal value of the decoupled SDP on the right hand side (RHS) of (\ref{eq:3.3}), that is,
\begin{displaymath}
	\mathrm{SDP}(\mathbf{X})= \min_{(\mathbf{u}_x, \mathbf{u}_y) \in S_{\mathbf{X}}^+(\mathbf{u}_x, \mathbf{u}_y) } g(\mathbf{u}_x, \mathbf{u}_y).
\end{displaymath}

Next we show that $\mathrm{SDP}(\mathbf{X})=\|\mathbf{X}\|_{\mathcal{A}_M}$ by proving: {\em i)}  $\mathrm{SDP}(\mathbf{X})\leq \|\mathbf{X}\|_{\mathcal{A}_M}$, and {\em ii)}  $\mathrm{SDP}(\mathbf{X})\geq \|\mathbf{X}\|_{\mathcal{A}_M}$. 

\subsection{First, we show i) $\mathrm{SDP}(\mathbf{X})\leq \|\mathbf{X}\|_{\mathcal{A}_M}$.} 
\begin{lemma}
	\label{lm:a1}
	For $\forall \mathbf{X}\in\mathbb{C}^{N\times M}$, $\exists (\tilde{\mathbf{u}}_x, \tilde{\mathbf{u}}_y) \in S_X^+$ such that
	\begin{displaymath}
		g(\tilde{\mathbf{u}}_x, \tilde{\mathbf{u}}_y) = \|\mathbf{X}\|_{\mathcal{A}_M}.
	\end{displaymath} 
\end{lemma}

Proof: According to Theorem \ref{th:1}, for any $\mathbf{X}\in\mathbb{C}^{N\times M}$, there exist a unique decomposition
\begin{displaymath}
	\mathbf{X}=\sum_l s_l \mathbf{A}(\mathbf{f}_l), \quad \mathbf{A}(\mathbf{f}_l)\in\mathcal{A}_M
\end{displaymath}
such that
\begin{displaymath}
	\|\mathbf{X}\|_{\mathcal{A}_M}=\sum_l |s_l|.
\end{displaymath}

Now we construct $\mathbf{T}(\tilde{\mathbf{u}}_x)$ and $\mathbf{T}(\tilde{\mathbf{u}}_y)$ as follows:
\begin{equation}
	\label{eq:t1}
	\mathbf{T}(\tilde{\mathbf{u}}_x)=\textstyle \sum_l \sqrt{\frac{M}{N}} |s_l| \mathbf{a}_N(f_{x, l}) \mathbf{a}_N^\mathrm{H}(f_{x, l}),
\end{equation}
and
\begin{equation}
	\label{eq:t2}
	\mathbf{T}(\tilde{\mathbf{u}}_y)=\textstyle \sum_l  \sqrt{\frac{N}{M}} |s_l| \mathbf{a}_M(f_{y, l}) \mathbf{a}_M^\mathrm{H}(f_{y, l}).
\end{equation}

We can easily verify that
\begin{equation}
	\begin{split}
		&\left(\begin{array}{cc}
			\mathbf{T}(\mathbf{u}_y) & \mathbf{X}^\mathrm{H} \\
			\mathbf{X} & \mathbf{T}(\mathbf{u}_x)
		\end{array}\right)\\
		&=\sum_l \frac{1}{\sqrt{NM}} |s_l| \left(\begin{array}{c}
			\sqrt{N}\mathbf{a}_M(f_{y, l}) \\
			\mathrm{sign}(s_l)\sqrt{M}\mathbf{a}_N(f_{x, l}) 
		\end{array}\right) \\
		&\cdot\left(\begin{array}{c}
			\sqrt{N}\mathbf{a}_M(f_{y, l}) \\
			\mathrm{sign}(s_l)\sqrt{M}\mathbf{a}_N(f_{x, l})
		\end{array}\right)^\mathrm{H} \succeq \mathbf{0}.
	\end{split}
\end{equation}
That is,  $(\tilde{\mathbf{u}}_x, \tilde{\mathbf{u}}_y) \in S_X^+$. 

Thus,
\begin{equation}
	\begin{split}
		g(\tilde{\mathbf{u}}_x, \tilde{\mathbf{u}}_y) &=\frac{1}{2\sqrt{NM}}\bigg(\mathrm{trace}\big(\mathbf{T}(\mathbf{u}_x)\big)+\mathrm{trace}\big(\mathbf{T}(\mathbf{u}_y)\big)\bigg) \\
			&=\textstyle \sum_l |s_l| \\
			&=\|\mathbf{X}\|_{\mathcal{A}_M}.
	\end{split}
\end{equation}
\QED

\begin{corollary}
	$\mathrm{SDP}(\mathbf{X})\leq \|\mathbf{X}\|_{\mathcal{A}_M}$ for $\forall \mathbf{X}\in\mathbb{C}^{N\times M}$.
\end{corollary}

Proof: This is straightforward from Lemma A.1, since $\mathrm{SDP}(\mathbf{X})$ is the minimum of $g(\cdot)$. \QED 

%
%
%
%

\subsection{Next, we show ii) $\mathrm{SDP}(\mathbf{X})\geq \|\mathbf{X}\|_{\mathcal{A}_M}$. }
\label{sec:app2.2}

We make a key observation that the problem of 2-D linear spectrum estimation from a single snapshot can be alternatively viewed as a constrained 1-D problem (say along $x$ dimension) of recovering $\mathbf{f}_x$ from multiple measurement vectors (MMV). A general 1-D MMV problem is studied in  \cite{yang2014exact}, which defines the following MMV atom set:
\begin{equation}
		\mathcal{A}_x=\left\{\mathbf{a}_N(f)\mathbf{e}_M^H, \ \forall f\in[0, 1],  \ \forall \mathbf{e}_M\in\mathbb{C}^{M\times 1}: \|\mathbf{e}_M\|=1\right\}. \label{eq:Ax-set}
	\end{equation}
Essentially, each atom in $\mathcal{A}_x$ is an $N\times M$ matrix whose columns consist of $M$ weighted copies of $\mathbf{a}_N(f)$, where the weighting vector $\mathbf{e}_M$ is normalized to have unit length. 

For MMV problems, the following results in \cite{yang2014exact} are useful. 

\begin{lemma}{[Theorem 3 (MMV SDP)] \cite{yang2014exact}.}
	\label{lm:a3}
	For any $\mathbf{X}\in\mathbb{C}^{N\times M}$ that can be linearly decomposed over the MMV atom set in (\ref{eq:Ax-set}), 
	its atomic norm over $\mathcal{A}_x$ can be calculated via the following SDP (denoted as $\mathrm{SDP}_x(\mathbf{X})$) 
	\begin{equation}
		\label{eq:a5}
		\begin{split}
			\|\mathbf{X}\|_{\mathcal{A}_x}=&\min_{\mathbf{V}, \mathbf{u}_x} \left\{\frac{1}{2\sqrt{N}}\bigg(\mathrm{trace}(\mathbf{V})+\mathrm{trace}\big(\mathbf{T}(\hat{\mathbf{u}}_x)\big)\bigg)\right. \\
			&  ~~~~ \mathrm{s.t.}~ \left(\begin{array}{cc}
				\mathbf{V} & {\mathbf{X}}^\mathrm{H} \\
				{\mathbf{X}} & \mathbf{T}({\mathbf{u}}_x)
			\end{array}\right)\succeq \mathbf{0},
		\end{split}
	\end{equation}
	where $\mathbf{V}\in\mathbb{C}^{M\times M}$ is some Hermitian matrix and $\mathbf{T}(\hat{\mathbf{u}}_x)\in\mathbb{C}^{N\times N}$ is Toeplitz.
\end{lemma}

\begin{lemma}{[Theorem 4 (MMV ANM)] \cite{yang2014exact}.}
	\label{lm:a4}
	If $\mathbf{X}=\sum_l c_l \mathbf{A}_x(f_{x, l})$, $\mathbf{A}_x(f_{x, l})\in\mathcal{A}_x$, satisfies the frequency separation condition
	\begin{equation}
		\label{eq:a22}
		\Delta_{\min, x}=\min_{i\neq j} |f_{x, i}-f_{x, j}|\geq\frac{1}{\lfloor(N-1)/4\rfloor},
	\end{equation}
	then it is guaranteed that
	\begin{equation}
		\|\mathbf{X}\|_{\mathcal{A}_x}=\sum_l |c_l|. \label{eq:a4.2}
	\end{equation}
\end{lemma}

With these two lemmas, we are ready to prove ii) as follows. 

\begin{lemma}
	\label{lm:a2}
	Suppose that $\mathbf{X}=\sum_l s_l \mathbf{A}(\mathbf{f}_l)$ where $\mathbf{A}(\mathbf{f}_l)\in\mathcal{A}_M$ satisfies the sufficient separation condition (\ref{eq:2.5}). Then, it hold that 
	\begin{equation}
		\mathrm{SDP}(\mathbf{X})
		\geq  \|\mathbf{X}\|_{\mathcal{A}_M}.
	\end{equation} 
\end{lemma}

Proof: Without losing of generality, we inspect the $x$ dimension and assume
\begin{displaymath}
	\Delta_{\min, x}\geq\frac{1}{\lfloor(N-1)/4\rfloor}.
\end{displaymath}

When $\mathbf{X}$ can be linearly decomposed by the matrix-form atom set $\mathcal{A}_M$, it also allows for a linear decomposition over $\mathcal{A}_x$, as follows: 
\begin{equation}
	\begin{split}
		\mathbf{X}&=\sum_l s_l \mathbf{a}_N(f_{x, l})\mathbf{a}_M^\mathrm{H}(f_{y, l}) \\
			&=\sum_l s_l \|\mathbf{a}_M^\mathrm{H}(f_{y, l})\| \mathbf{a}_N(f_{x, l})\frac{\mathbf{a}_M^\mathrm{H}(f_{y, l})}{\|\mathbf{a}_M^\mathrm{H}(f_{y, l})\|} \\
			&=\sum_l \left(\sqrt{M} s_l \right) \mathbf{a}_N(f_{x, l})\frac{\mathbf{a}_M^\mathrm{H}(f_{y, l})}{\|\mathbf{a}_M^\mathrm{H}(f_{y, l})\|},
	\end{split} \label{eq:a4-c}
\end{equation}
where $\|\mathbf{a}_M^\mathrm{H}(f)\|=\sqrt{M}, \forall f\in[0, 1]$, $\frac{\mathbf{a}_M^\mathrm{H}(f_{y, l})}{\|\mathbf{a}_M^\mathrm{H}(f_{y, l})\|}$ has unit length, and hence $\mathbf{a}_N(f_{x, l})\frac{\mathbf{a}_M^\mathrm{H}(f_{y, l})}{\|\mathbf{a}_M^\mathrm{H}(f_{y, l})\|} \in \mathcal{A}_x$.

According to Lemma \ref{lm:a4}, $\mathbf{X}$ satisfies (\ref{eq:a22}), which means $\mathbf{f}_x$ can be revealed via the MMV atomic norm minimization as in (\ref{eq:a4.2})  \cite{yang2014exact}. Specifically, Lemma \ref{lm:a4} and (\ref{eq:a4-c}) lead to 
\begin{displaymath}
	\|\mathbf{X}\|_{\mathcal{A}_x}=\sqrt{M} \sum_l |s_l|.
\end{displaymath}

Meanwhile, Theorem \ref{th:1} suggests
\begin{displaymath}
	\|\mathbf{X}\|_{\mathcal{A}_M}=\sum_l |s_l|.
	\end{displaymath}
Hence, we reach the following equality: 
\begin{equation}
	\label{eq:a40}
	\|\mathbf{X}\|_{\mathcal{A}_M}=\frac{1}{\sqrt{M}} \|\mathbf{X}\|_{\mathcal{A}_x}.
\end{equation}

Next,  comparison between  (\ref{eq:3.3}) and  (\ref{eq:a5}) reveals that the 2-D problem formulated in $\mathrm{SDP}(\mathbf{X})$ shares the same objective function as  $\frac{1}{\sqrt{M}}\mathrm{SDP}_x(\mathbf{X})$ for the 1-D MMV problem, but the former has an extra constraint that $\mathbf{T}(\tilde{\mathbf{u}}_x)$ (or $\mathbf{V}$ in (\ref{eq:a5})) is Topelitz. Since, the minimal point $\mathrm{SDP}(\mathbf{X})$ is a constrained version of $\frac{1}{\sqrt{M}}\mathrm{SDP}_x(\mathbf{X})$, we  have
\begin{equation}
	\label{eq:a41}
	\mathrm{SDP}(\mathbf{X})\geq \frac{1}{\sqrt{M}}\mathrm{SDP}_x(\mathbf{X})=\frac{1}{\sqrt{M}}\|\mathbf{X}\|_{\mathcal{A}_x}.
\end{equation}

Putting together (\ref{eq:a40}) and (\ref{eq:a41}), we conclude
\begin{equation}
	\label{eq:a8}
	\mathrm{SDP}(\mathbf{X})\geq\|\mathbf{X}\|_{\mathcal{A}_{M}}.
\end{equation}

\QED


}

\ifCLASSOPTIONcaptionsoff
  \newpage
\fi

\bibliographystyle{IEEEtran}
\bibliography{IEEEabrv,overview}

\end{document}